\def\be{\begin{equation}}
\def\ee{\end{equation}}
\def\ba{\begin{array}}
\def\ea{\end{array}}

\documentclass[preprint,showpacs,amsmath]{revtex4}
\usepackage{amsfonts}
\usepackage{amssymb}
\def\qed{\leavevmode\unskip\penalty9999 \hbox{}\nobreak\hfill
     \quad\hbox{\leavevmode  \hbox to.77778em{%
               \hfil\vrule   \vbox to.675em%
               {\hrule width.6em\vfil\hrule}\vrule\hfil}}
     \par\vskip3pt}

\newtheorem{theorem}{Theorem}

\input amssym.def
\begin{document}
\title{\large\bf Trade-off relations on CHSH tests for multi-qubit pure states }
\author{ Meiyu Cui$^{1,2}$, Xiaofen Huang$^{1}$,  Tinggui Zhang$^{1,\dag}$\\[10pt]
\footnotesize
\small$1$ School of Mathematics and Statistics, Hainan Normal University,\\
\small Haikou 571158, P. R. China\\
\small$2$ Center School of Anji Vocational Education, Huzhou,
313301,
P. R. China  \\
\small $^\dag$ Correspondence to tinggui333@163.com}

\bigskip

\begin{abstract}

We study the trade-off relations on the maximal violation of CHSH
tests for the multi-qubit pure states. Firstly, according to the
classification of 3-qubit pure states under stochastic local
operations and classical communication, four trade-off relations on
CHSH tests are provided. The process of proof also provides the
method of calculating its exact value. Then we consider the
multi-qubit quantum systems, and we prove that the conjecture in
[PRA 94, 042105 (2016)] is right for some special 4-qubit or more
qubit pure states.
\end{abstract}

 \maketitle

\bigskip
\section{Introduction }

Quantum entanglement as one of the most fascinating features of
quantum mechanics has been investigated for decades. The relation of
quantum entanglement and the violation of Bell's inequalities
\cite{aa,bb,cc} is one of the key issues in quantum information. And
the existence of Bell inequalities \cite{dd} and their observed
violation in experiments have had a very deep impact on the way we
look at quantum mechanics. Quantum mechanics exhibits the
nonlocality of nature by violation of the Bell inequality.

Entanglement and nonlocality both are essential
resource in quantum information theory, and establishing
the link between them is intriguing.
 Quantum entanglement is considered to be
the most nonclassical manifestation of quantum mechanics. It is exciting to
 know which states are both entangled and nonlocal. Quantum entanglement
 coincides with the violation of Bell inequalities for pure quantum states.
 Any pure entangled state violate a Bell inequality, and all bipartite pure entangled state violate the Bell inequality
  and the magnitude of the violation is directly
  proportional to the amount entanglement of the states \cite{b,c,d,e,f,g,szha}. The necessary and sufficient condition for a 2-qubit mixed state to violate the
  Bell-CHSH(Clauser-Horrne-Shimony-Holt) inequality has been derived \cite{h}. A key property of entanglement is that a quantum system entangled with one other
   limits its entanglement with the remaining ones. This phenomenon is known as quantum monogamy and has recently been widely studied \cite {vjw,ndsv,j,scmj}. In particular,
   the authors of \cite{j} presented the analytical trade-off relations obeyed by the CHSH test of pairwise qubits in a
    3-qubit system.

 Two states have the same kind of entanglement if they can be obtained with certainty from each other via local operation and classical communication
 (LOCC) with nonzero probability. In \cite{k}, $\rm {W. D\ddot{u}r}$ et al. presented that SLOCC splits the set of pure
states of 3-qubit into six inequivalent classes.

\textbf{Class A-B-C(product states)}
\begin{eqnarray}
|\psi_{A-B-C}\rangle=|0\rangle|0\rangle|0\rangle,
\end{eqnarray}
where the rank of the reduced density matrices of $|\psi_{A-B-C}\rangle$ satisfies $r(\rho_A)=r(\rho_B)=r(\rho_C)=1$.

\textbf{Class A-BC,AB-C and C-AB(bipartite entanglement states)}
\begin{eqnarray}
|\psi_{A-BC}\rangle=|0\rangle(C_\delta|0\rangle|0\rangle+S_\delta|1\rangle|1\rangle),
\end{eqnarray}
 where we denote $Sin{\delta}, Cos{\delta}$  with $S_{\delta},C_{\delta}$ respectively in abbreviation in this paper, and $C_\delta\geq S_\delta>0$, $r(\rho_A)=1, r(\rho_B)=r(\rho_C)=2$, and similarly for $|\psi_{B-AC}\rangle$ and $|\psi_{C-AB}\rangle$.

\textbf{W-class}
\begin{eqnarray}
|\psi_W\rangle=\sqrt{a}|001\rangle+\sqrt{b}|010\rangle+\sqrt{c}|100\rangle+\sqrt{d}|000\rangle,
\end{eqnarray}
where $a, b, c>0$, $d=1-(a+b+c)\geq0.$

\textbf{GHZ-class}
\begin{eqnarray}
|\psi_{GHZ}\rangle=\sqrt{k}(C_\delta|0\rangle|0\rangle|0\rangle+S_\delta e^{i\varphi}|\psi_A\rangle|\psi_B\rangle|\psi_C\rangle),
\end{eqnarray}
where $|\psi_A\rangle=C_\alpha|0\rangle+S_\alpha|1\rangle$, $|\psi_B\rangle=C_\beta|0\rangle+S_\beta|1\rangle$,
$|\psi_C\rangle=C_\gamma|0\rangle+S_\gamma|1\rangle$, and $\kappa=(1+2C_\delta S_\delta C_\alpha C_\beta C_\gamma C_\varphi)^{-1}\in(\frac{1}{2},+\infty)$
is a normalization factor, the ranges of five parameters are $\delta\in (0,\frac{\pi}{4}], \alpha,\beta,\gamma\in (0,\frac{\pi}{2}]$ and $\varphi\in(0,2\pi]$.


Now we consider the 3-qubit quantum states via the violation of
Bell inequality by the two-part reduced density matrix of the
3-qubit quantum states.

For a 2-qubit state $\rho$, it can be expressed according to the Bloch representation, that is
\begin{eqnarray}
\rho=\frac{1}{4}(I\otimes I+\sum_{i=1}^3{r_i\sigma_i \otimes I}+\sum_{j=1}^3{s_j I\otimes\sigma_j}+
\sum_{i,j=1}^3{m_{ij} \sigma_i\otimes\sigma_j}),
\end{eqnarray}
where $I$ is the identity matrix, $\sigma_i$, $i=1,2,3$ are the Pauli matrices, and the coefficients
$r_i={\rm tr}(\rho\sigma_i\otimes I)$, $s_j={\rm tr}(\rho I\otimes \sigma_j)$, the correlation matrix $ m_{ij}={\rm tr}(\rho\sigma_i\otimes\sigma_j)$,
 and $M=(m_{kl})$ is a matrix with size $3\times 3$.

The famous CHSH inequality was derived  in 1969 \cite{dd}
$$
|\langle CHSH\rangle|\leq 2,
$$
where CHSH operator is $A_1B_1+A_1B_2+A_2B_1-A_2B_2$, $A_1, A_2$ are observables on Alice, $B_1, B_2$ are observables on Bob.
Combining the Bloch representation, one has the CHSH relation for bipartite quantum state $\rho$ given in \cite{l}£¬
 \begin{eqnarray}
 \langle CHSH\rangle_\rho=2\sqrt{\tau_1+\tau_2},
 \end{eqnarray}
 where $\tau_1$ and $\tau_2$ are the two largest eigenvalues of the matrix $M^\dag M$.

Later Qin et al. presented a trade-off relation about CHSH inequality for the 3-qubit
states in \cite{j}. For any 3-qubit state $\rho_{ABC}$ in quantum system $\mathbf{H^A}\otimes\mathbf{H^B}\otimes\mathbf{H^C\mathrm{}}$, the maximal violation of CHSH tests on pairwise bipartite states satisfies the following trade-off relation:
\begin{eqnarray}\label{eq1}
\langle CHSH\rangle_{\rho_{AB}}^2+\langle
CHSH\rangle_{\rho_{BC}}^2+\langle CHSH\rangle_{\rho_{AC}}^2\leq12,
\end{eqnarray}
where $\rho_{AB}={\rm tr}_C\rho_{ABC}$, $\rho_{AC}={\rm
tr}_B\rho_{ABC}$, $\rho_{BC}={\rm tr}_A\rho_{ABC}$.

In this paper we derive four CHSH trade-off relations based on the
six classes of 3-qubit quantum pure states.

\section{Specific trade-off relations for 3-qubit pure states}
According to the six classes of 3-qubit pure stats, we can obtain different trade-off relations respectively.
\begin{theorem}
 For the 3-qubit product states(class A-B-C), the maximal violation of CHSH inequality tests on pairwise bipartite states satisfies the following trade-off relation
\begin{eqnarray}\label{re1}
\langle CHSH\rangle_{\rho_{AB}}^2+\langle CHSH\rangle_{\rho_{BC}}^2+\langle CHSH\rangle_{\rho_{AC}}^2=12.
\end{eqnarray}
\end{theorem}

{\bf{Proof:}} Let the 3-qubit product pure state be
$|\psi\rangle=|0\rangle|0\rangle|0\rangle$, which has density matrix
 $\rho_{ABC}=|\psi\rangle\langle\psi|=|000\rangle\langle000|$.

So the reduced density matrices are
$\rho_{BC}=\rho_{AC}=\rho_{AB}=|00\rangle\langle00|$,
and they are in the 2-qubit quantum subsystem.

Now we decomposite the reduced density matrix $\rho_{AB}$ in Bloch representation, and we can get the entries of correlation matrix are
\begin{eqnarray*}
&m_{11}^{AB}=m_{12}^{AB}=m_{13}^{AB}=m_{21}^{AB}=m_{22}^{AB}=m_{23}^{AB}=m_{31}^{AB}=m_{32}^{AB}=0,\\
&m_{33}^{AB}=1.
\end{eqnarray*}
So the correlation matrix $M^{AB}$ is
\begin{eqnarray*}
M^{AB}=
\left(
 \begin{array}{ccc}
  0&0&0\\
  0&0&0\\
  0&0&1
 \end{array}
 \right),
\end{eqnarray*}
and the eigenvalues of matrix ${M^{AB}}^\dag M^{AB}$ are $\tau_1=1$
and $\tau_2=\tau_3=0$.

Then one get the mean value of  CHSH operator for the reduced matrices $\rho_{AB}$, $\rho_{BC}$, $\rho_{AC}$
\begin{eqnarray*}
\langle CHSH\rangle_{\rho_{AB}}^2=\langle CHSH\rangle_{\rho_{BC}}^2=\langle CHSH\rangle_{\rho_{AC}}^2=4(\tau_1+\tau_3)=4.
\end{eqnarray*}
Thus we can obtain the trade-off CHSH relation (\ref{re1})  for the 3-qubit product states(class A-B-C) in the theorem. $\Box$

\begin{theorem}
For bipartite entanglement states(class A-BC,AB-C,C-AB) $\rho_{ABC}$, it satisfies the following trade-off relation
\begin{eqnarray}\label{re2}
8\leq\langle CHSH\rangle_{\rho_{AB}}^2+\langle CHSH\rangle_{\rho_{BC}}^2+\langle CHSH\rangle_{\rho_{AC}}^2<12.
\end{eqnarray}
\end{theorem}
{\bf{Proof:}}
Without loss of generality , let the bipartite entanglement states (class A-BC) be $|\psi_{A-BC}\rangle=|0\rangle(C_\delta|0\rangle|0\rangle+S_\delta|1\rangle|1\rangle)$, and it is with density matrix $\rho_{ABC}$,
\begin{eqnarray*}
\rho_{ABC}=C_{\delta}^2|000\rangle\langle000|+S_{\delta} C_{\delta}|011\rangle\langle000|+C_{\delta} S_{\delta}|000\rangle\langle011|+S_{\delta}^2|011\rangle\langle011|,
\end{eqnarray*}
where $\quad C_{\delta}\geq S_{\delta}>0$,
and the ranks of reduced matrices satidfy $r(\rho_A)=1$, $r(\rho_B)=r(\rho_C)=2$.

Taking trace on the subsystem $\mathbf{H^C\mathrm{}}$, one has the reduced matrix $\rho_{AB}$ like this
\begin{eqnarray*}
\rho_{AB}=C_{\delta}^2|00\rangle\langle00|+S_{\delta}^2|01\rangle\langle01|,
\end{eqnarray*}
and we rewrite $\rho_{AB}$ with matrix form
\begin{eqnarray*}
 \rho_{AB}=\left(
 \begin{array}{cccc}
  C_{\delta}^2&0&0&0\\
  0&S_{\delta}^2&0&0\\
  0&0&0&0\\
  0&0&0&0\\
 \end{array}
 \right).
\end{eqnarray*}
Then denoting $\rho_{AB}$ in Bloch representation, we can get the correlation matrix $M^{AB}$ of $\rho_{AB}$
\begin{eqnarray*}
M^{AB}=
\left(
 \begin{array}{ccc}
  0&0&0\\
  0&0&0\\
  0&0&C_{\delta}^2-S_{\delta}^2
 \end{array}
 \right).
\end{eqnarray*}
Since the matrix $({M^{AB}})^\dag M^{AB}$ has the eigenvalues
$\tau_1=[C_{\delta}^2-S_{\delta}^2]^2$ and $\tau_2=\tau_3=0$, we
arrive at that
\begin{eqnarray}\label{b1}
\langle CHSH\rangle_{\rho_{AB}}^2=4(\tau_1+\tau_3)=4(C_{\delta}^2-S_{\delta}^2)^2.
\end{eqnarray}

Similarly we get equalities about CHSH equalities of reduced matrices $\rho_{BC}$ and $\rho_{AC}$
\begin{equation}
\begin{aligned}\label{b2}
&\langle CHSH\rangle_{\rho_{BC}}^2=16C_{\delta}^2S_{\delta}^2+4,\\
&\langle CHSH\rangle_{\rho_{AC}}^2=4(C_{\delta}^2-S_{\delta}^2)^2.
\end{aligned}
\end{equation}
Combining equalities (\ref{b1}) and (\ref{b2}),  we have
\begin{equation}\label{a}
\begin{aligned}
 &\langle CHSH\rangle_{\rho_{AB}}^2+\langle CHSH\rangle_{\rho_{BC}}^2+\langle CHSH\rangle_{\rho_{AC}}^2\\
 =&8(C_{\delta}^2-S_{\delta}^2)^2+16C_{\delta}^2S_{\delta}^2+4 =8C_{2\delta}^2+4S_{2\delta}^2+4=4C_{2\delta}^2+8.\\
\end{aligned}
\end{equation}
Because of $C_{\delta} > 0$ and $C_{2\delta}=1-2C_{\delta}^2<1$, thus $0\leq C_{2\delta}^2<1$. Also due to the range of ${C_{2\delta}}^2$ and Eq.(\ref{a}), then we can get that the trade-off relation (\ref{re2}). $\Box$

The following conclusion is about the W-class state.
\begin{theorem}
For the W-class state $\rho_{ABC}$, there is the following trade-off relation
\begin{eqnarray}\label{re3}
8 < \langle CHSH\rangle_{\rho_{AB}}^2+\langle
CHSH\rangle_{\rho_{BC}}^2+\langle CHSH\rangle_{\rho_{AC}}^2 < 12.
\end{eqnarray}
\end{theorem}
{\bf{Proof:}} Let the 3-qubit pure state of W-class be
$
|\psi_W\rangle=\sqrt{a}|001\rangle+\sqrt{b}|010\rangle+\sqrt{c}|100\rangle+\sqrt{d}|000\rangle,
$
where $a, b, c>0$, $d=1-(a+b+c)\geq0.$

By calculating the eigenvalues of correlation matrix of
$|\psi_W\rangle$ and borrowing the results of Ref.\cite{ppam} , we
can obtain the CHSH relations in the flowing
\begin{eqnarray*}
\begin{aligned}
&\langle CHSH\rangle_{\rho_{AB}}^2=2[1+12ab-4ac-4bc+\sqrt{V}],\\
&\langle CHSH\rangle_{\rho_{AC}}^2=2[1+12ac-4ab-4bc+\sqrt{V}],\\
&\langle CHSH\rangle_{\rho_{BC}}^2=2[1+12bc-4ab-4ac+\sqrt{V}],
\end{aligned}
\end{eqnarray*}
where
$V=[(\sqrt{a}+\sqrt{b}+\sqrt{c})^2+d][(\sqrt{a}+\sqrt{b}-\sqrt{c})^2+d][(\sqrt{a}-\sqrt{b}+\sqrt{c})^2+d][(-\sqrt{a}+\sqrt{b}+\sqrt{c})^2+d]$.
Therefore
\begin{equation}
\begin{aligned}
&\langle CHSH\rangle_{\rho_{AB}}^2+\langle CHSH\rangle_{\rho_{BC}}^2+\langle CHSH\rangle_{\rho_{AC}}^2\\
=&2[3+4ab+4ac+4bc+3\sqrt{V}]=2[3(1+\sqrt{V})+4(ab+ac+bc)].
\end{aligned}
\end{equation}
We consider the above formula as a function of parameters $a, b$ and
$c$, then it's a continuous function of parameters $ a, b$ and $ c$.
Thus, when $ a, b$ and $ c$ all tend to zero, the value of this
function approaches $12$. When $a = b$ approaches $1/2$, the value
of the function approaches $8$. In fact, for any  3-qubit pure
state, the value of the function will be between $8$ and $12$. The
upper bound has been proved \cite{j}, and the lower bound is
obvious. Because the sum of all eigenvalues of $
M_{AB}^{\dag}M_{AB}$, $M_{AC}^{\dag}M_{AC}$and $M_{BC}^{\dag}M_{BC}$
is $3$, and here we use the largest two eigenvalues of each matrix,
so this value will not be less than $3\times 2/3=2$, that is
$\langle CHSH\rangle_{\rho_{AB}}^2+\langle
CHSH\rangle_{\rho_{BC}}^2+\langle CHSH\rangle_{\rho_{AC}}^2\geq
4\times 2=8$. Therefore, we get the trade-off relation (\ref{re3}).
$\Box$

\begin{theorem}
For the GHZ-class state $\rho_{ABC}$, there is the following trade-off relation
\begin{eqnarray}\label{re4}
8\leq\langle CHSH\rangle_{\rho_{AB}}^2+\langle CHSH\rangle_{\rho_{BC}}^2+\langle CHSH\rangle_{\rho_{AC}}^2\leq12.
\end{eqnarray}
\end{theorem}
{\bf{Proof:}} Let the GHZ-class pure state be with form
\begin{eqnarray}
|\psi_{GHZ}\rangle=\sqrt{k}(C_\delta|0\rangle|0\rangle|0\rangle+S_\delta e^{i\varphi}|\psi_A\rangle|\psi_B\rangle|\psi_C\rangle),
\end{eqnarray}
where $|\psi_A\rangle=C_\alpha|0\rangle+S_\alpha|1\rangle$, $|\psi_B\rangle=C_\beta|0\rangle+S_\beta|1\rangle$,
$|\psi_C\rangle=C_\gamma|0\rangle+S_\gamma|1\rangle$, and $\kappa=(1+2C_\delta S_\delta C_\alpha C_\beta C_\gamma C_\varphi)^{-1}\in(\frac{1}{2},+\infty)$
is a normalization factor, the ranges of five parameters are $\delta\in (0,\frac{\pi}{4}], \alpha,\beta,\gamma\in (0,\frac{\pi}{2}]$ and $\varphi\in(0,2\pi]$.

Via computing the eigenvalues of reduced density matrix of $|\psi_{GHZ}\rangle$, and we can get CHSH relations in the flowing
\begin{eqnarray*}
\langle CHSH\rangle_{\rho_{AB}}^2=4\times[1+\frac{(C_\alpha^2-C_\beta^2-C_\gamma^2+2C_\beta^2C_\gamma^2)S_{2\delta}^2
-C_\alpha^2C_\beta^2C_\gamma^2S_{2\delta}^2}{(1+C_\alpha^2C_\beta^2C_\gamma^2S_{2\delta}^2)^2}],\\
\langle CHSH\rangle_{\rho_{AC}}^2=4\times[1+\frac{(C_\beta^2-C_\alpha^2-C_\gamma^2+2C_\alpha^2C_\gamma^2)S_{2\delta}^2
-C_\alpha^2C_\beta^2C_\gamma^2S_{2\delta}^2}{(1+C_\alpha^2C_\beta^2C_\gamma^2S_{2\delta}^2)^2}],\\
\langle CHSH\rangle_{\rho_{BC}}^2=4\times[1+\frac{(C_\gamma^2-C_\alpha^2-C_\beta^2+2C_\alpha^2C_\beta^2)S_{2\delta}^2
-C_\alpha^2C_\beta^2C_\gamma^2S_{2\delta}^2}{(1+C_\alpha^2C_\beta^2C_\gamma^2S_{2\delta}^2)^2}].
\end{eqnarray*}
therefore
\begin{gather}\label{f1}
\begin{aligned}
&\langle CHSH\rangle_{\rho_{AB}}^2+\langle CHSH\rangle_{\rho_{AC}}^2+CHSH\rangle_{\rho_{BC}}^2\\
=&12+4\times\frac{(2C_\alpha^2C_\beta^2+2C_\alpha^2C_\gamma^2+2C_\beta^2C_\gamma^2-C_\alpha^2-C_\beta^2-C_\gamma^2
-3C_\alpha^2C_\beta^2C_\gamma^2)S_{2\delta}^2}
{(1+C_\alpha^2C_\beta^2C_\gamma^2S_{2\delta}^2)^2}.
\end{aligned}
\end{gather}

Let $a=C_\alpha^2, b=C_\beta^2, c=C_\gamma^2$, where $a,b,c\in[0,1)$, we consider a function about variables $a, b, c$
\begin{gather}
\begin{aligned}
f(a,b,c)=&2C_\alpha^2C_\beta^2+2C_\alpha^2C_\gamma^2+2C_\beta^2C_\gamma^2-C_\alpha^2-C_\beta^2-C_\gamma^2
-3C_\alpha^2C_\beta^2C_\gamma^2\\
=&2ab+2ac+2bc-a-b-c-3abc\\
=&a(b-1)+b(c-1)+c(a-1)+ab(1-c)+ac(1-b)+bc(1-a)\\
=&a(b-1)(1-c)+b(c-1)(1-a)+c(a-1)(1-b)\leq 0\\
\end{aligned}
\end{gather}
At the same time, we can get the lower bound of $f(a,b,c)$ is $-1$, i. e.
\begin{equation}\label{f2}
-1\leq f(a,b,c)\leq 0,
\end{equation}

and also
$0< S_{2\delta}^2\leq1$, $(1+C_\alpha^2C_\beta^2C_\gamma^2S_{2\delta}^2)^2>1$,
thus we have
\begin{eqnarray}\label{f3}
0<\frac{S_{2\delta}^2}{(1+C_\alpha^2C_\beta^2C_\gamma^2S_{2\delta}^2)^2}<1.
\end{eqnarray}
 Combining (\ref{f1}), (\ref{f2}), (\ref{f3}), we can obtain the trade-off relation (\ref{re4}). $\Box$

{\bf Remark:} In fact, if we only look at the conclusions of our
theorems, we will find that after a lot of calculations we only get
similar results. This makes our conclusions look a little weak. But
we must emphasize that the really useful conclusions are the
formulas $(10), (14)$ and $(17)$ in the theorems proof, because they
give the exact values of each specific quantum state.

A brief summary, for all six classes of the 3-qubit pure states, the
quantity $\langle CHSH\rangle_{\rho_{AB}}^2+\langle
CHSH\rangle_{\rho_{BC}}^2+\langle CHSH\rangle_{\rho_{AC}}^2$
relative to CHSH operators been calculated accurately and bound
between 8 and 12.

\section{Trade-off relations for multi-qubit pure states}
Now, we study trade-off relations about CHSH for multi-qubit pure states.  For all of 4-qubit
pure states $|\psi_{ABCD}\rangle$, the authors of the paper\cite{scam}
conjectured that the following inequality holds
$$tr[T_{AB}T_{AB}^{T}]+tr[T_{AC}T_{AC}^{T}]+tr[T_{AD}T_{AD}^{T}] \leq
3,$$ with respect to the bi-partitions (AB,CD),(AC,BD),and (AD,BC),
and $T_{AB}=tr[\rho_{ABCD}\sigma_i\otimes\sigma_j\otimes I_2\otimes
I_2]$, $\rho_{ABCD}=|\psi_{ABCD}\rangle\langle\psi_{ABCD}|$ and so
on. Through the discussion of the three-qubit pure state in the
previous section, we found that the conjecture is correct in at
least the following two situations.

(1) When $|\psi_{ABCD}\rangle$ is a fully separable state. Then by
local unitary transformation $|\psi_{ABCD}\rangle=|0000\rangle$,
therefore,
$Tr[T_{AB}T_{AB}^{T}]=Tr[T_{AC}T_{AC}^{T}]=Tr[T_{AD}T_{AD}^{T}]=1$,
i.e
$Tr[T_{AB}T_{AB}^{T}]+Tr[T_{AC}T_{AC}^{T}]+Tr[T_{AD}T_{AD}^{T}]=3.$

(2) When $|\psi_{ABCD}\rangle$ is a generalized GHZ state. That is
to say
$|\psi_{ABCD}\rangle=\cos\theta|0000\rangle+\sin\theta|1111\rangle$.
It is easy to get the following facts
$Tr[T_{AB}T_{AB}^{T}]=Tr[T_{AC}T_{AC}^{T}]=Tr[T_{AD}T_{AD}^{T}]=\cos\theta^2+\sin\theta^2=1$,
i.e
$Tr[T_{AB}T_{AB}^{T}]+Tr[T_{AC}T_{AC}^{T}]+Tr[T_{AD}T_{AD}^{T}]=3.$

If like the three-qubit state, only the fully separable state and
the GHZ state can give the upper bound for Eq.(\ref{eq1}), then the
conjecture of \cite{scam} is right.  The conclusions of this kind of
conjecture under the special case can be extended to more qubit
cases.
\section{ Conclusions and discussions}
In this paper, we have calculated in detail the restriction
conditions of CHSH inequality violation values for different classes
of three qubit quantum state. We have also studied the 4-qubit
situation. These enables us to have a deeper understanding of
monogamy relations of three and more qubit pure quantum states.

\bigskip
\noindent{\bf Acknowledgments}\  \  This project is
 supported by the National Natural Science Foundation of China Grants No.11861031 and the Natural Science Foundation of Hainan province under Grants No.118QN230.

\end{document}